\documentclass[aps,twocolumn,showpacs]{revtex4}
\usepackage{graphicx}
\usepackage{amssymb}
\usepackage{amsmath}

\def\ba{\begin{eqnarray}}
\def\ea{\end{eqnarray}}
\def\be{\begin{equation}}
\def\ee{\end{equation}}

\def\bm{\begin{math}}
\def\me{\end{math}}

\def\la{\langle}
\def\ra{\rangle}

\def\pra#1#2#3{Phys. Rev. A \textbf{#1}, #2 (#3)}
\def\prb#1#2#3{Phys. Rev. B \textbf{#1}, #2 (#3)}
\def\pre#1#2#3{Phys. Rev. E \textbf{#1}, #2 (#3)}

\begin{document}

\title{Domain Growth in Ising Systems with Quenched Disorder}

\thispagestyle{empty}

\author{Raja Paul{$^1$}, Sanjay Puri$^2$ and Heiko Rieger{$^1$}}

\affiliation{$^1$Theoretische Physik, Universit\"at des Saarlandes,
  66041 Saarbr\"ucken, GERMANY}

\affiliation{$^2$School of Physical Sciences, Jawaharlal Nehru
  University, New Delhi -- 110067, INDIA.}

\begin{abstract}
We present results from extensive Monte Carlo (MC) simulations of domain growth
in ferromagnets and binary mixtures with quenched disorder.
These are modeled by the {\it random-bond Ising model} and the
{\it dilute Ising model} with either nonconserved (Glauber)
spin-flip kinetics or conserved (Kawasaki) spin-exchange kinetics. In all
cases, our MC results are consistent with power-law growth with an exponent
$\theta (T,\epsilon)$ which depends on the quench temperature $T$ and the
disorder amplitude $\epsilon$. Such exponents arise naturally when the
coarsening domains are trapped by energy barriers which grow logarithmically
with the domain size. Our MC results show excellent agreement with the
predicted dependence of $\theta (T,\epsilon)$.

\end{abstract}


\maketitle

\section{Introduction}

Consider a binary mixture which is homogeneous at high temperatures. This
system becomes thermodynamically unstable if it is quenched below the critical
temperature. The subsequent evolution of the system
is characterized by the formation and
growth of domains enriched in either component. These domains have a
characteristic size $R(t)$, which grows with time. The domain growth law
[$R(t)$ vs. $t$] depends on general system properties, e.g., the nature
of conservation laws governing the order parameter evolution; the presence
of hydrodynamic velocity fields; the presence of quenched or annealed disorder,
etc. There is a good understanding of the growth laws for pure and isotropic
systems \cite{ab94,bf01,ao02,dp03}.
For the case with nonconserved order parameter, e.g., ordering of a
magnet into up and down phases, the system obeys the Lifshitz-Cahn-Allen (LCA)
growth law, $R(t) \sim t^{1/2}$. For the case with conserved order parameter,
e.g., diffusion-driven phase separation of an AB mixture into A-rich and B-rich
phases, the system obeys the Lifshitz-Slyozov (LS) growth law, $R(t) \sim
t^{1/3}$.

Recent interest in domain growth problems
has focused on modeling and understanding the
effects of various experimentally relevant features. In this context,
an important set of analytical and numerical studies has investigated coarsening
in systems with quenched disorder
\cite{hh85,gs85,oc86,cgg87,pcp91,pp92,hh91,bh91,ghs95}.
In general, one expects that
trapping of domain boundaries by disorder sites will result in slower domain
growth. However, these studies were unable to clarify the nature (or even
existence) of a universal growth law. In a recent letter \cite{ppr04},
we have revisited this problem through comprehensive Monte Carlo (MC)
simulations of kinetic Ising models. In our letter, we presented MC results for
ordering in random magnets, modeled by the {\it random-bond
Ising model} (RBIM) with nonconserved (Glauber) spin-flip kinetics.
(In the RBIM, the presence of disorder is
mimicked by randomizing the exchange coupling between spins.) In this paper,
we present further results for coarsening in two classes of disordered systems: \\
(a) The RBIM with conserved (Kawasaki) spin-exchange kinetics, which models
phase separation in disordered binary mixtures. \\
(b) The {\it dilute Ising model} (DIM) with both nonconserved
and conserved kinetics. The DIM is relevant in cases where disorder is introduced
via either {\it bond dilution} or {\it site dilution}.

The results in this paper, in conjunction with those in our letter, constitute a
novel understanding of domain growth in systems with quenched disorder.
This paper is organized as follows. In Sec.~II, we summarize arguments for
growth laws in disordered systems. In Sec.~III, we present results for the RBIM
with conserved kinetics. In Sec.~IV, we present results for the DIM
with both nonconserved and conserved kinetics. Finally, Sec.~V
concludes this paper with a summary and discussion of our results.

\section{Growth Laws in Disordered Systems}

\subsection{Nonconserved Case}

An important step towards understanding growth laws
in nonconserved systems is due to Lai et al. (LMV) \cite{lmv88}. LMV
proposed four classes of systems, determined by the
dependence of the energy barrier to coarsening on the
characteristic scale. The growth of domains is driven by a
curvature-reduction mechanism as
\ba
\label{curv}
{dR \over dt} = {a(R,T) \over R} ,
\ea
where the diffusion constant $a(R,T)$ depends on the domain
scale $R$ and temperature $T$, in general. For pure systems, the diffusion
constant is independent of the length scale, i.e., $a(R,T) = a_0$. The
corresponding growth law is the LCA law, $R(t) = (2 a_0 t)^{1/2}$.

Let us next consider systems with quenched disorder. At early times and small
length-scales, the growing domains are not affected by disorder
[$a(R,T) \simeq a_0$] and the growth law is the same as that for the pure case.
At late times, the domains are trapped by disorder sites,
creating a barrier ($E_B$) to domain growth. Then, the asymptotic
dynamics is driven by thermal activation over
disorder barriers with $a(R,t) \simeq a_0 \exp (-\beta E_B)$, where
$\beta = T^{-1}$ ($k_B = 1$). For the bond-disordered case,
Huse and Henley (HH) \cite{hh85} argued that the energy barrier scales as
$E_B(R) \simeq \epsilon R^{\psi}$, where $\epsilon$ is the disorder
strength. The barrier exponent $\psi$ depends on
the roughening exponent $\zeta$ and the pinning exponent $\chi$ as
$\psi = \chi/(2- \zeta)$. Further, the roughening and pinning
exponents are related as $\chi = 2 \zeta + d-3$, where $d$ is the
dimensionality. For power-law barriers, Eq.~(\ref{curv}) yields an
asymptotic growth law which is {\it logarithmic}, viz.,
\ba
\label{hh}
R(t) &\simeq& \left[ \frac{T}{\epsilon} \ln \left( \frac{t}{t_0} \right)
\right]^{1/\psi} , \nonumber \\
t_0 &\simeq& \frac{1}{a_0 \psi} \left( \frac{T}{\epsilon} \right)^{2/\psi} .
\ea
We can reformulate the early-time and late-time behaviors as limiting cases
of a crossover function:
\ba
\label{cross}
R(t) = R_0 (T,\epsilon) h\left( \frac{t}{t_0} \right),
\ea
where 
\ba
\label{r0}
R_0(T,\epsilon)= \left(\frac{T}{\epsilon}\right)^{1/\psi} ,
\ea
and
\ba
\label{hx}
h(x) &=& \left( \frac{2}{\psi} x \right)^{1/2}, \quad x \ll 1 , \nonumber \\
&=& \left( \ln x \right)^{1/\psi}, \quad x \gg 1 .
\ea

For $d=2$, $\zeta = 2/3$ and $\chi = 1/3$
\cite{fns77,hhf85}, yielding $\psi = 1/4$. For $d=3$, a perturbative
calculation gives $\psi \simeq 0.55$ \cite{hh85}.
There have been a number of numerical simulations
\cite{gs85,oc86,cgg87,pcp91,pp92,hh91,bh91,ghs95}
and experiments \cite{iei90,lla00,llop01} which have attempted to test the
HH scenario. However, to date, there is no clear confirmation of HH growth
in the asymptotic regime. As a matter of fact, it is not even clear whether
there is a universal law which characterizes the disorder-affected
growth regime.

In recent work \cite{ppr04}, we have revisited this problem via extensive MC
simulations of the RBIM with nonconserved kinetics. Our results
were consistent with power-law domain growth, but with a temperature- and
disorder-dependent exponent. Similar observations have been made in
experiments on coarsening in disordered systems \cite{iei90,lla00,llop01}.
Such growth exponents can be understood in the framework of a
logarithmic (rather than power-law) $R$-dependence of trapping barriers.
In the context of the DIM, Henley \cite{ch85} and Rammal and Benoit
\cite{rb85} have argued that the fractal nature of domain boundaries results
in a logarithmic $R$-dependence of energy barriers. We propose that this
is generally applicable and examine the implications thereof \cite{remark}.
Recall that, at early times and small length scales, we expect disorder-free
domain growth. Then, the appropriate logarithmic barrier-scaling form is as
follows:
\ba
\label{blog}
E_B(R) \simeq \epsilon \ln \left( 1 + R \right) ,
\ea
where $R$ is measured in dimensionless units. Substituting $a(R,T) \simeq a_0 \exp
(- \beta E_B)$ in Eq.~(\ref{curv}), we obtain
\begin{equation}
\label{log}
\frac{dR}{dt} = \frac{a_0}{R} (1+R)^{-\epsilon/T} .
\end{equation}
The solution of Eq.~(\ref{log}) is
\ba
\label{hrb}
R(t) &\simeq& (2 a_0 t)^{1/2}, \quad t \ll t_0, \nonumber \\
&\simeq& \left[ \left(2+ \frac{\epsilon}{T}\right)~a_0t
\right]^{\theta (T,\epsilon)}, \quad t \gg t_0 ,
\ea
with the asymptotic growth exponent
\ba
\label{ncexp}
\theta (T,\epsilon) = \frac{1}{2+\epsilon/T} .
\ea
The crossover length and time can be identified by rewriting Eq.~(\ref{hrb})
in the form of Eq.~(\ref{cross}) with
\ba
R_0 &=& {1 \over (2\theta)^{\theta/(1-2\theta)}}, \nonumber \\
t_0 &=& \frac{1}{a_0} {1 \over (2\theta^{2 \theta})^{1/(1-2 \theta)}},
\ea
and
\ba
h(x) &=& x^{1/2}, \quad x \ll 1, \nonumber \\
&=& x^{\theta}, \quad x \gg 1 .
\ea

In our letter \cite{ppr04}, we have shown that the growth exponent
for the nonconserved RBIM is consistent with Eq.~(\ref{ncexp}). Let us
next discuss the implications of power-law and logarithmic barriers
for domain growth with conserved kinetics.

\subsection{Conserved Case}

In the absence of disorder, the domain scale obeys the Huse equation
\cite{dh86}:
\ba
\label{huse}
{dR \over dt} = {D_0 \over R^2},
\ea
with the solution $R(t) = (3D_0 t)^{1/3}$. The presence of disorder
renormalizes the diffusion constant $D_0$ by an Arrhenius factor:
$D(R,T) \simeq D_0 \exp (- \beta E_B)$. For
logarithmic barriers as in Eq.~(\ref{blog}), the corresponding
growth equation is
\ba
\label{husel}
{dR \over dt} = {D_0 \over R^2} \left( 1 + R \right)^{-\epsilon/T} .
\ea
The short-time and long-time solutions of Eq.~(\ref{husel}) are obtained as
follows:
\ba
\label{cdis}
R(t) &\simeq& (3D_0 t)^{1/3}, \quad t \ll t_0, \nonumber \\
&\simeq& \left[ \left(3+ \frac{\epsilon}{T}\right) D_0 t \right]^{\theta (T,\epsilon)},
\quad t \gg t_0,
\ea
where
\ba
\label{cexp}
\theta (T,\epsilon) = {1 \over 3+\epsilon/T} .
\ea
The crossover form of Eq.~(\ref{cdis}) is Eq.~(\ref{cross}) with
\ba
R_0 &=& {1 \over (3\theta)^{\theta/(1-3\theta)}}, \nonumber \\
t_0 &=& {1 \over D_0} {1 \over (3\theta^{3\theta})^{1/(1-3\theta)}} ,
\ea
and
\ba
h(x) &=& x^{1/3}, \quad x \ll 1, \nonumber \\
&=& x^{\theta}, \quad x \gg 1.
\ea

Notice that the asymptotic exponent differs from that for the nonconserved case
when the energy-barriers are logarithmic. This should
be contrasted with the HH scenario, where the asymptotic growth law is the
same for the nonconserved and conserved cases \cite{pp92}. This is easily seen
by incorporating the HH barrier-scaling form in Eq.~(\ref{huse}).

\section{Random Bond Ising Model: Conserved Kinetics}

\subsection{Modeling and Numerical Details}

The Hamiltonian for the RBIM is as follows:
\begin{equation}
{\cal H} = -\sum_{\langle ij \rangle} J_{ij} S_i S_j, \quad S_i = \pm 1 .
\label{rbim}
\end{equation}
For a binary (AB) mixture, the spins $S_i$ label whether a lattice site $i$ is
occupied by an A-atom (say, $S_i=+1)$ or a B-atom $(S_i=-1)$. We consider the
case where the spins are placed on an $L^2$ square lattice with periodic
boundary conditions. We introduce quenched
disorder in the exchange coupling $J_{ij}$, corresponding to immobile
impurities in a binary mixture. The $J_{ij}$'s have a uniform
distribution on the interval $[1-\epsilon/2, 1+\epsilon/2]$, where
$\epsilon$ quantifies the amount of disorder. The limit $\epsilon =2$
corresponds to maximum disorder, and $\epsilon = 0$ corresponds to the pure
case. (We confine ourselves to the case where the exchange couplings are
always ferromagnetic, $J_{ij} \geq 0$.) The subscript
$\la ij \ra$ in Eq.~(\ref{rbim}) denotes a sum over nearest-neighbor pairs only.
For the pure case, $T_c^{\rm{pure}} \simeq 2.269$ for a $d=2$ square lattice.
Since the average coupling strength is $\langle J_{ij} \rangle = 1$,
as in the pure case, the critical
temperature remains almost unaltered, $T_c \in [2.0, 2.269]$ \cite{XX}. Assigning
random initial orientations to each spin, we rapidly quench the system
to $T<T_c$. The initial condition corresponds to a critical quench, with
50 \% A (up) and 50 \% B (down).

The Ising model has no intrinsic dynamics as the commutator of the spin
variables and the Hamiltonian is identically zero. Therefore, we introduce stochastic
dynamics by placing the system in contact with a heat bath. The resultant
dynamical model is referred to as a {\it kinetic Ising model}. The appropriate
stochastic kinetics for a binary mixture is Kawasaki spin-exchange
or conserved kinetics, where a randomly-selected spin $S_i$ is
exchanged with a randomly-chosen neighbor, $S_i \leftrightarrow S_j$. The
spin exchange is accepted with probability
\begin{eqnarray}
\label{tp}
W = \left\{
\begin{array}{cll}
\exp (- \beta \Delta {\cal H}) & \mathrm{for} &\Delta {\cal H} \ge 0 , \\
1 & \mathrm{for} &\Delta {\cal H} \le 0 ,
\end{array} \right.
\end{eqnarray}
where $\Delta \cal{H}$ is the change in energy resulting from the spin
exchange:
\ba
\label{deltah}
\Delta {\cal H} = (S_i-S_j) \left( \sum_{L_i \neq j} J_{iL_i} S_{L_i} -
\sum_{L_j \neq i} J_{jL_j} S_{L_j} \right) .
\ea
In Eq.~(\ref{deltah}), $L_i$ refers to the nearest-neighbors of lattice site
$i$. A single Monte Carlo step (MCS) corresponds to attempted updates of $L^2$
spins. A naive implementation of the Kawasaki model is numerically
demanding, and it has proven notoriously difficult to access the asymptotic LS
growth regime in the pure case \cite{asm88,mb95}. A number of accelerated
algorithms have been proposed in the literature \cite{mc_barkema} --
we employ the so-called {\it continuous-time algorithm}. In this approach,
a list of oppositely-oriented spins is prepared from the lattice configuration.
Then, a pair is selected randomly from the list, and is exchanged
according to Eq.~(\ref{tp}). In each trial, time is advanced by
$\Delta t = 1/n_t$, where $n_t$ is the total number of anti-aligned spin
pairs at time $t$. After each exchange, the list is updated.
This algorithm works particularly efficiently at low temperatures, where
bulk domains are strongly enriched in one component.

The segregating system is usually characterized by studying the time-dependence
of the correlation function:
\begin{equation}
\label{eq3}
C({\vec r},t)= \frac{1}{N} \sum_{i=1}^N
\left[ \langle S_i(t) S_{i+\vec{r}}(t) \rangle - \langle S_i(t) \rangle
\langle S_{i+\vec{r}}(t) \rangle \right]_{\rm{av}} ,
\end{equation}
which measures the overlap of the spin configuration at distance
$\vec{r}$. Here, $[...]_{\rm{av}}$ indicates an average
over different realizations of the bond disorder, and $\langle... \rangle$
denotes a thermal average, i.e., an average over different initial
configurations and realizations of the thermal noise. Typically, the growth
process is isotropic and characterized by a unique length scale $R(t)$. In
that case, the correlation function has a dynamical-scaling
form \cite{bs74}:
\begin{equation}
C(\vec{r},t) = g \left( \frac{r}{R} \right),
\label{ds}
\end{equation}
where $g(x)$ is the scaling function.

The characteristic size $R(t)$ is defined from the correlation
function as the distance over which it decays to (say) zero or
half its maximum value. There are a number of different definitions of the
length scale, but these are all equivalent in the scaling regime. Subsequently,
we will present results for the correlation function and the domain growth law.

\subsection{Numerical Results}

\begin{figure}[htb]
\centering
\includegraphics[width=\linewidth]{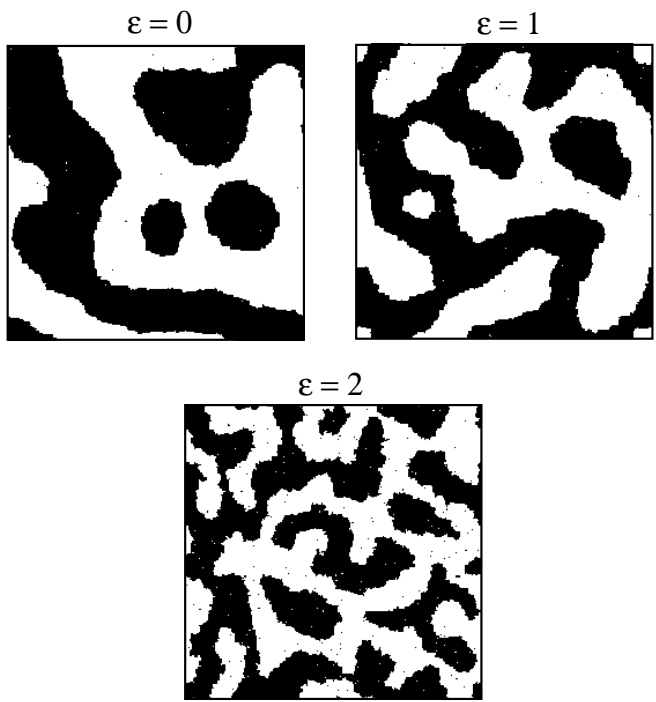}
\caption{Domain growth in the RBIM with Kawasaki kinetics. We show evolution
pictures at $t=10^7$ MCS for a $256^2$ lattice, after a quench
from $T = \infty$ to $T = 1.0$. The mixture has a critical composition
with 50 \% A ($S_i = +1$, marked in black) and 50 \% B ($S_i = -1$, unmarked).
The snapshots correspond to different disorder amplitudes: $\epsilon = 0$
(pure case), and $\epsilon = 1, 2$.}
\label{fig1}
\end{figure}

In Fig.~\ref{fig1}, we show evolution pictures for the conserved RBIM after
a critical quench from $T = \infty$ to $T = 1.0$. We show snapshots
at $t=10^7$ MCS for $\epsilon = 0$ (pure case), and $\epsilon = 1, 2$.
The domains have been identified by calculating the time-average for each
spin:
\begin{equation}
\label{eq8}
m_i = {\frac{1}{\Delta}}\sum_{t=t_i}^{t_f}S_i(t) ,
\end{equation}
within a suitable time-window $\Delta = t_f - t_i$. It is clear from
the snapshots that the evolution is slower for higher amplitudes
of disorder. This will be quantified via the corresponding domain growth laws.

Next, we consider the scaled correlation-function data [$C(r,t)$ vs. $r/R$]
for the morphologies in Fig.~\ref{fig1}. Our statistical
data for the RBIM is obtained on $d=2$ lattices of size $512^2$ (with
$T=1.0$ and $\epsilon$ being varied), and $256^2$ (with $\epsilon=2$
and $T$ being varied).
In order to improve the statistics, we averaged within a finite time-window
around each data point. Further, the data was obtained as an
average over 32 independent initial conditions for both the spin and
disorder configurations. The length scale $R$ is defined as the first
zero-crossing of the correlation function. 
We have confirmed that $C(r,t)$ exhibits
dynamical scaling [as in Eq.~(\ref{ds})]
for different disorder amplitudes and quenches to different values of
$T$.

\begin{figure}[htb]
\centering
\includegraphics[width=\linewidth]{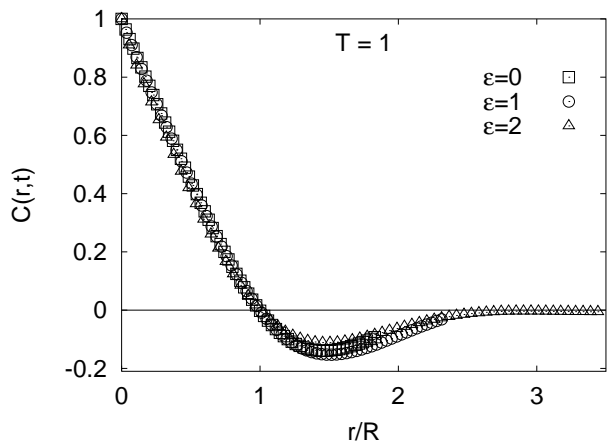}
\caption{Scaling plot of the correlation function for the evolution
depicted in Fig.~\ref{fig1}. We plot $C(r,t)$ vs.
$r/R$ at $t=10^7$ MCS for disorder amplitudes $\epsilon = 0$ (pure case),
and $\epsilon = 1, 2$. The length scale is defined as the first zero-crossing
of $C(r,t)$.}
\label{fig2}
\end{figure}

In Fig.~\ref{fig2}, we show that the scaling
function is independent of the disorder amplitude. This has also been
demonstrated in earlier studies of phase 
separation in disordered systems \cite{pp92,ghs95}.
In physical terms, the universality of the scaling function means that the
morphologies are equivalent, regardless of the disorder amplitude. (This was
already suggested by the snapshots in Fig.~\ref{fig1}.)
The typical transverse displacement
of interfaces due to disorder roughening is $L^{\zeta/(2-\zeta)}$,
where $\zeta$ is the roughening exponent \cite{hh85}.
At late times, one has $L \gg L^{\zeta/(2-\zeta)}$, because $\zeta < 1$
above the lower critical dimension. (If $\zeta > 1$, disorder-induced
roughening would destroy long-range order in the system.) Thus,
in the asymptotic regime, the roughness is irrelevant compared to the
domain size. Therefore, the evolution morphologies and their
statistical properties should be independent of disorder at late times.

Next, let us investigate the time-dependence of the domain size.
First, we study $R(t)$ vs. $t$ for quenches to different temperatures.

\begin{figure}[htb]
\centering
\includegraphics[width=\linewidth]{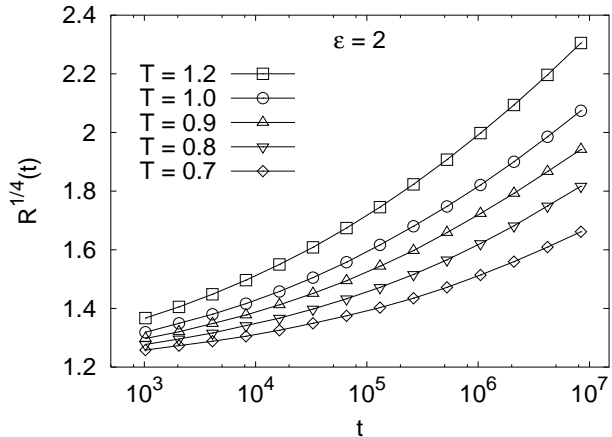}
\caption{Plot of $R^{1/4}$ vs. $t$ (on a log-linear scale) for $\epsilon = 2$
and different quench temperatures: $T = 0.7,0.8,0.9,1.0,1.2$.}
\label{fig3}
\end{figure}
In Fig.~\ref{fig3}, we undertake a direct test of the HH growth law in
Eq.~(\ref{hh}) by 
plotting $R^{1/4}$ vs. $\ln t$ for $\epsilon =2$ and different $T$-values.
Recall that $\psi = 1/4$ in $d=2$ according to the HH
argument, and the corresponding plot in Fig.~\ref{fig3} should
be linear in the asymptotic regime. However, the plot
exhibits continuous curvature and is not consistent with the HH growth law.
We have also attempted to fit the data to the functional form $\ln t=aR^x+b$. In
general, this function does not give a reasonable fit to the data. Even for
these poor fits, the exponent $x$ is strongly dependent on the temperature, at
variance with the prediction of a universal growth law. A similar observation
has been made in the experiments of Ikeda et al. \cite{iei90}, though these were
performed on random magnets, rather than disordered mixtures. As a matter of
fact, Ikeda et al. and Likodimos et al. \cite{lla00,llop01}
have argued that their experimental data for domain growth in disordered systems
is described by a power-law with a  temperature-dependent exponent
rather than the HH growth law. We have made a similar observation in our MC
studies of the nonconserved RBIM \cite{ppr04}. Let us examine the length-scale
data for the conserved RBIM from this perspective.
\begin{figure}[htb]
\centering
\includegraphics[width=\linewidth]{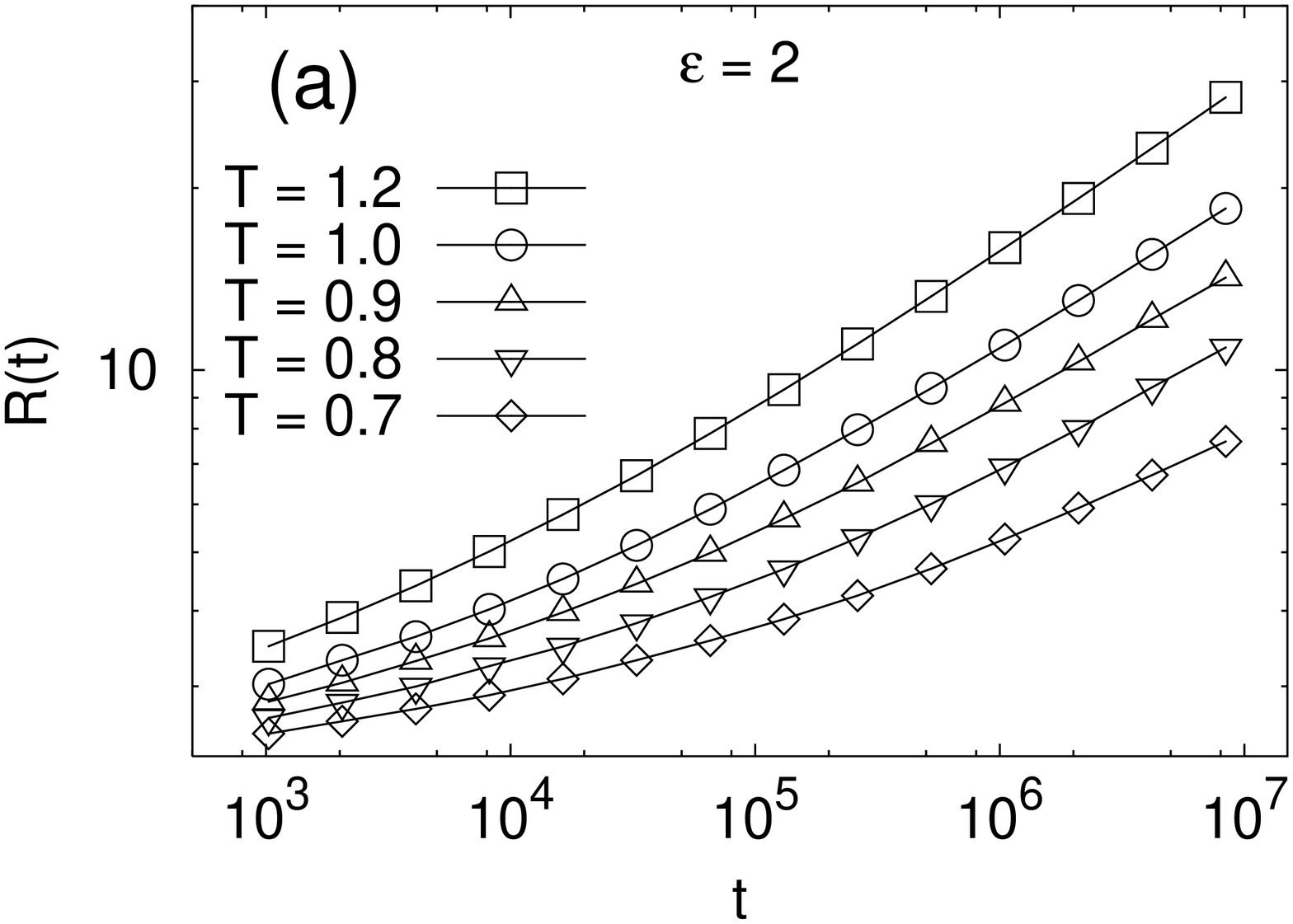}
\includegraphics[width=\linewidth]{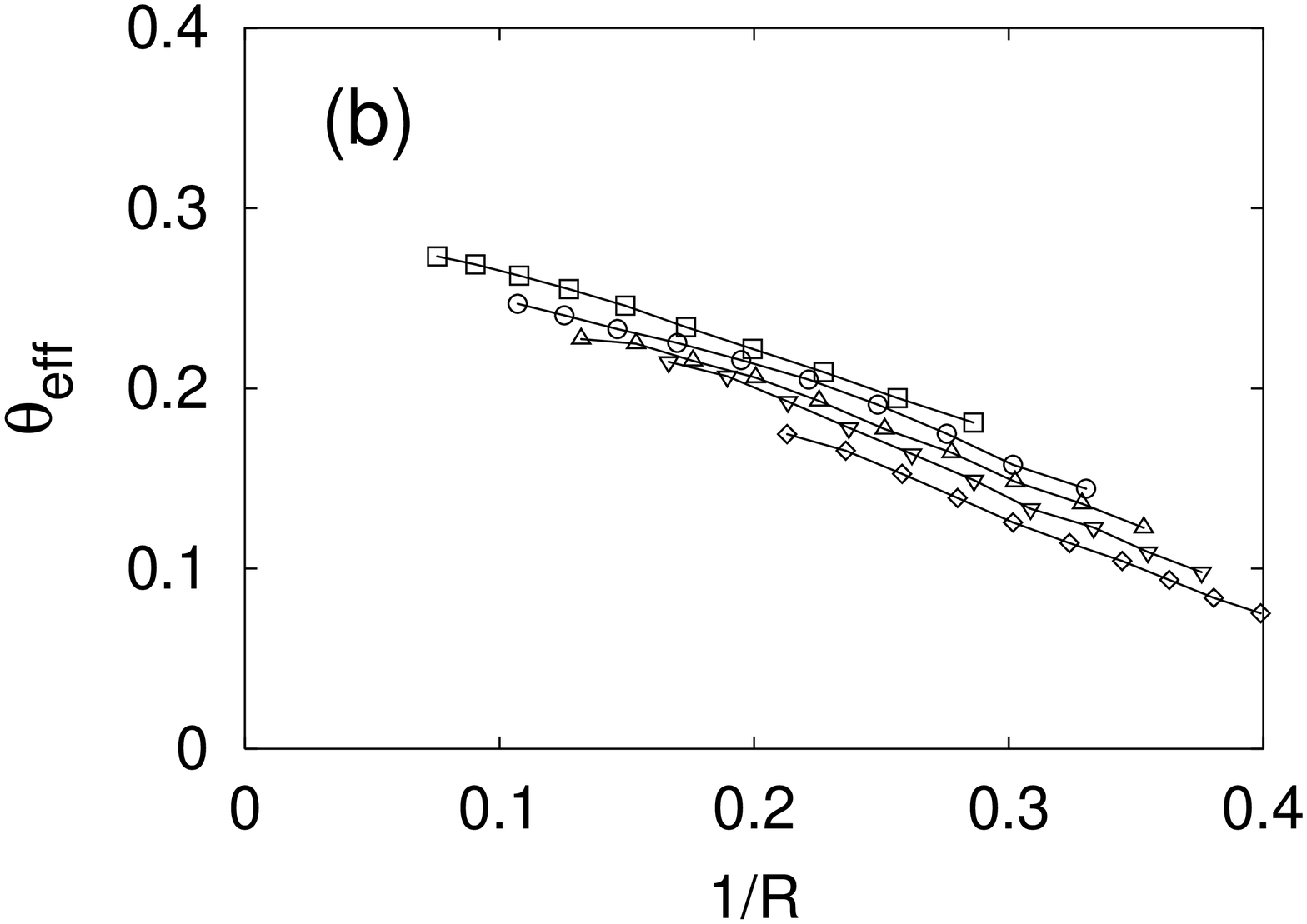}
\caption{(a) Plot of $R$ vs. $t$ (on a log-log scale) for the length-scale
data shown in Fig.~\ref{fig3}. (b) Plot of $\theta_{\rm{eff}} = d(\ln R)/d(\ln t)$
vs. $R^{-1}$ for the data in (a).}
\label{fig4}
\end{figure}

In Fig.~\ref{fig4}(a), we plot $R$ vs. $t$ from Fig.~\ref{fig3}
on a log-log scale. This plot does not show an extended linear
regime on the time-scale of our simulation. However, it is known that there
is an extended pre-asymptotic growth regime in the conserved case without
disorder \cite{dh86,asm88,mb95}, which complicates the observation of the LS growth
regime in MC simulations. Further, the slight upward curvature in the log-log plot
suggests that the growth law cannot be slower than a power law, at
variance with the HH result. In the pure case, Huse \cite{dh86} has
suggested that the asymptotic exponent may be obtained by extrapolating
the graph of the 
effective exponent $\theta_{\rm{eff}} = d(\ln R)/d(\ln t)$ vs. $R^{-1}$. We
apply a similar technique to the disordered case, and query whether the
resultant exponents are consistent with the scenario in Sec.~II.B
[cf. Eq.~(\ref{cexp})]. In Fig.~\ref{fig4}(b), we plot $\theta_{\rm{eff}}$ vs.
$R^{-1}$ for the data in Fig.~\ref{fig4}(a). The plots in Fig.~\ref{fig4}(b)
can be smoothly extrapolated to $R^{-1} = 0$ ($R = \infty$) to determine
$\theta = \theta_{\rm{eff}} (\infty)$, which depends on $T$.

\begin{figure}[htb]
\centering
\includegraphics[width=\linewidth]{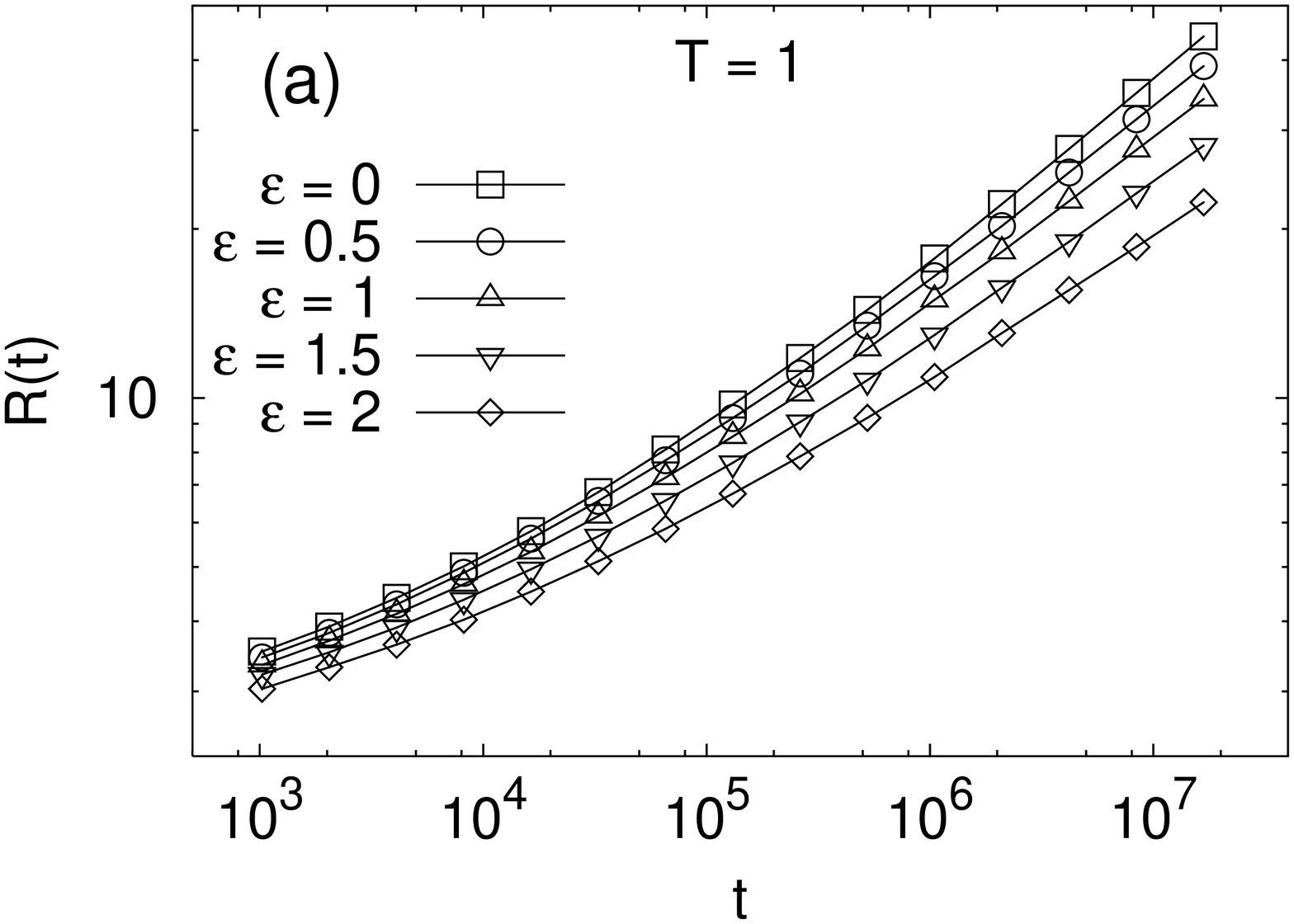}
\includegraphics[width=\linewidth]{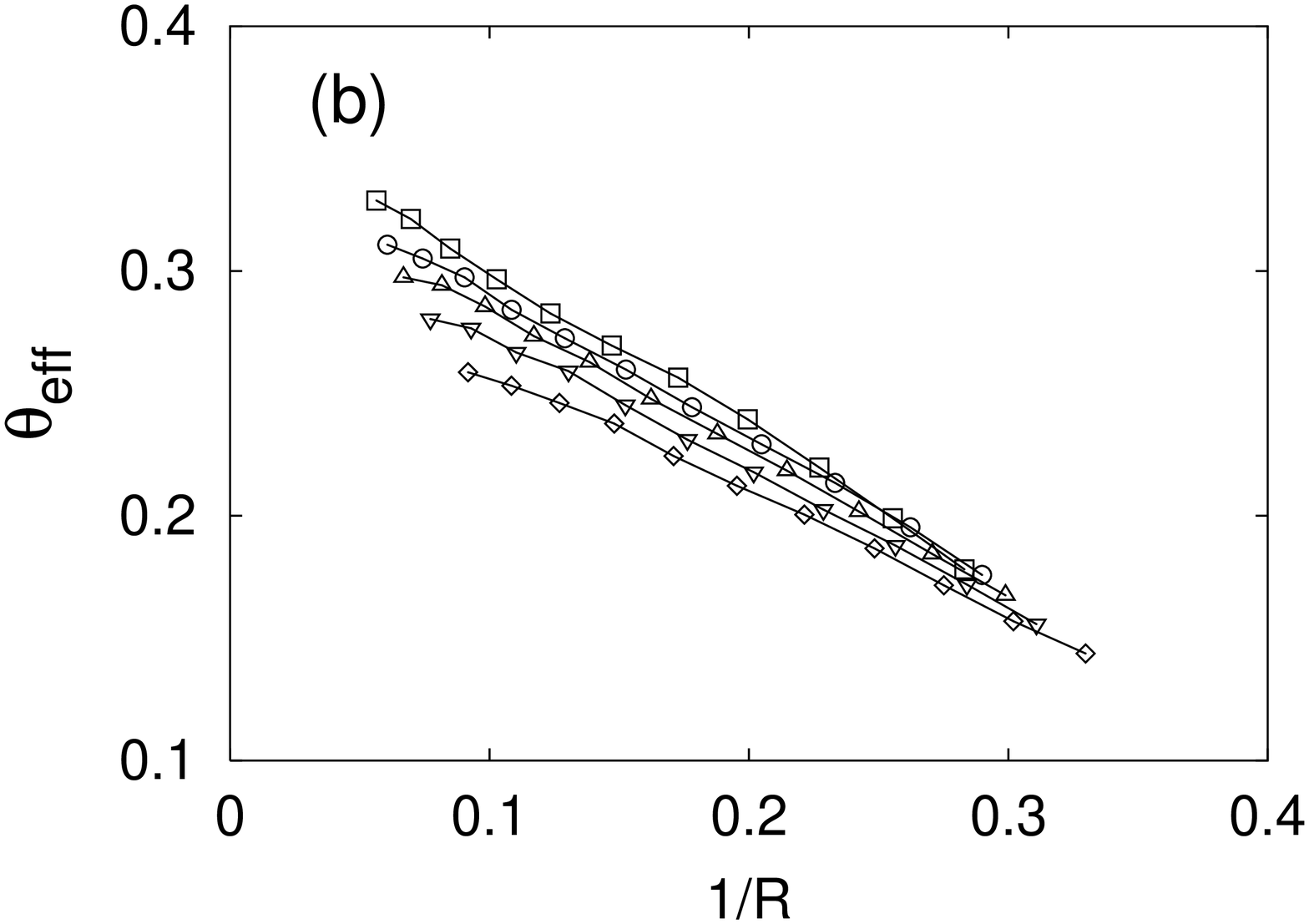}
\caption{(a) Plot of $R$ vs. $t$ (on a log-log scale) for $T=1.0$ and
different disorder amplitudes: $\epsilon = 0$ (pure case), and $\epsilon =
0.5,1.0,1.5,2.0$. (b) Plot of $\theta_{\rm{eff}} = d(\ln R)/d(\ln t)$
vs. $R^{-1}$ for the data in (a).}
\label{fig5}
\end{figure}
Next, we consider $R$ vs. $t$ at fixed temperature as the disorder amplitude
is varied. Again, we find that our data is not consistent with either the
HH scenario or even logarithmic growth. In Fig.~\ref{fig5}(a), we plot
$R$ vs. $t$ on a log-log scale for different $\epsilon$-values. The
corresponding plots of $\theta_{\rm{eff}}$ vs. $R^{-1}$ are shown in
Fig.~\ref{fig5}(b). In this case, the asymptotic exponent depends on the
disorder amplitude. Notice that we have also shown data for the pure case
($\epsilon = 0$) in Fig.~\ref{fig5}(a). We do not
see an extended linear regime even in this case.
However, the corresponding plot of $\theta_{\rm{eff}}$ vs. $R^{-1}$ in
Fig.~\ref{fig5}(b) extrapolates to the well-known LS value, $\theta \simeq 0.33$.

In Sec.~II.B, we have seen that a logarithmic barrier-scaling results in power-law
growth with varying exponents. We would like to test whether the asymptotic
exponents are consistent with the result in Eq.~(\ref{cexp}).
\begin{figure}[htb]
\centering
\includegraphics[width=\linewidth]{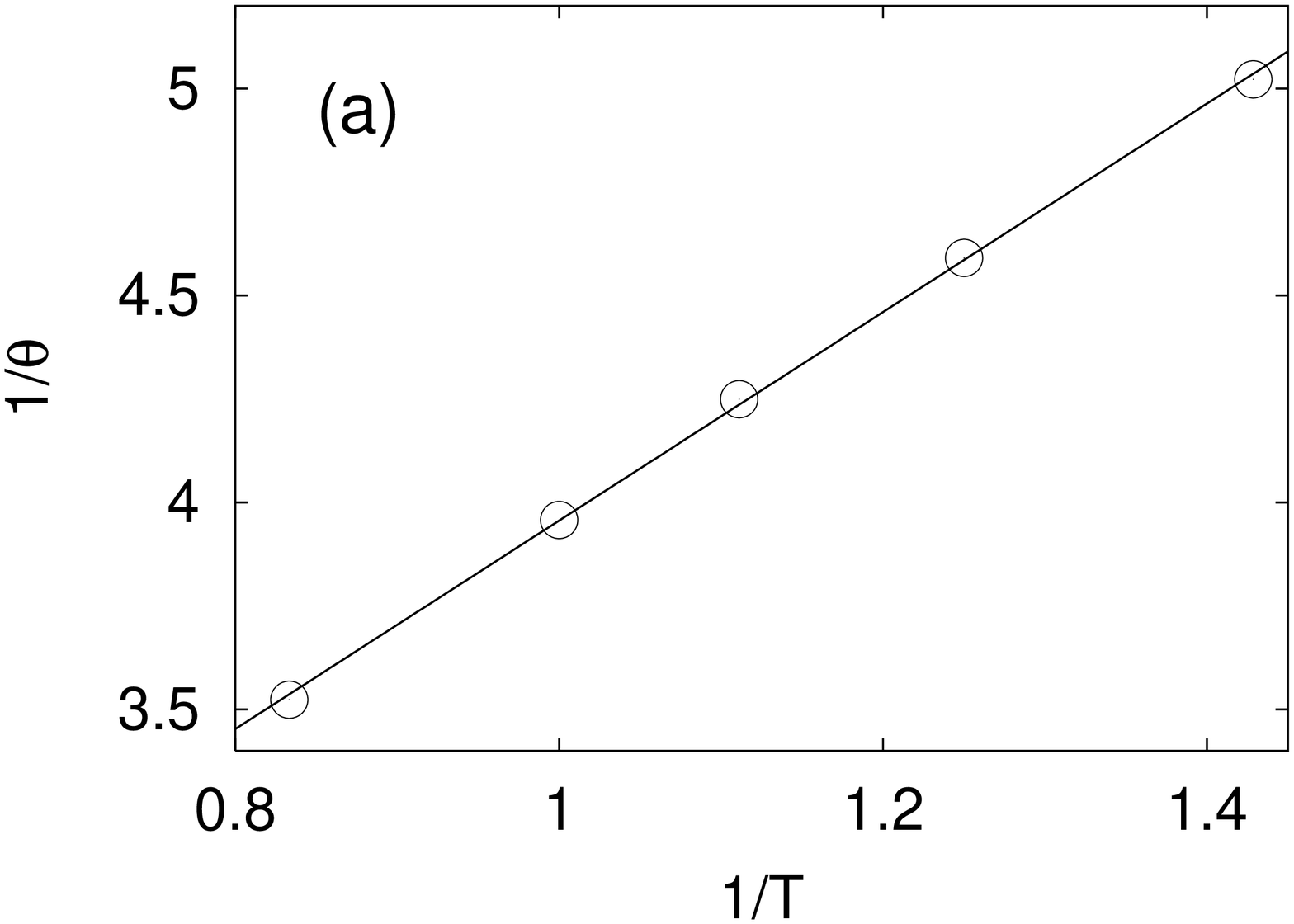}
\includegraphics[width=\linewidth]{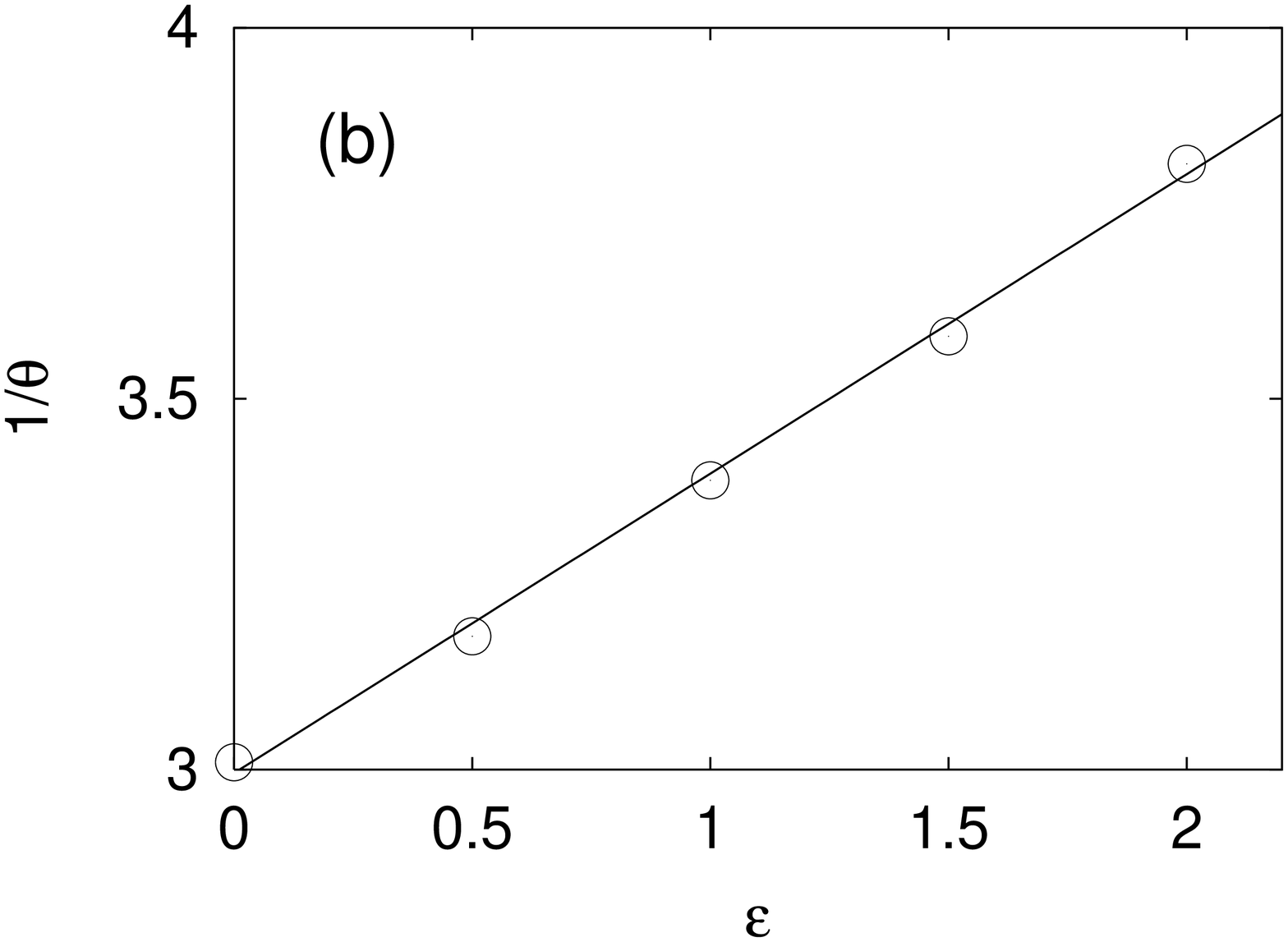}
\caption{(a) Exponent $1/\theta$ vs. $1/T$ for the data in Fig.~\ref{fig4}.
The solid line denotes the best linear fit to the data. (b) Exponent $1/\theta$
vs. $\epsilon$ for the data in Fig.~\ref{fig5}.}
\label{fig6}
\end{figure}
In Figs.~\ref{fig6}(a)
and (b), we plot $\theta^{-1}$ vs. $T^{-1}$ and $\epsilon$, respectively.
The resultant linear plots strongly support the logarithmic barrier-scaling scenario.

\section{Dilute Ising Model}

\subsection{Modeling and Numerical Details}

Next, we turn our attention to the DIM, where bond disorder is introduced by
diluting the spins on the lattice. The corresponding Hamiltonian is
\begin{equation}
{\mathcal{H}} =
-J\sum_{\langle ij \rangle}\varepsilon_i \varepsilon_j  S_i S_j , \quad
S_i = \pm 1 ,
\label{dham}
\end{equation}
with $J>0$. In Eq.~(\ref{dham}), the
$\varepsilon_i$'s are quenched, uncorrelated random variables with the
probability distribution:
\ba
P(\varepsilon) = p \delta_{\varepsilon,1} + (1-p) \delta_{\varepsilon,0} .
\ea
For a ferromagnet, $\varepsilon_i=0$ implies that
the magnetic atom at $i$ is replaced by
a non-magnetic impurity. In the context of an AB mixture, $\varepsilon_i=0$
corresponds to an immobile (non-interacting) impurity at site $i$.
Thus, there is no exchange interaction between the
atom at site $i$ and its nearest neighbors.
The distinguishing feature of the DIM (in contrast to the RBIM discussed in
Sec.~III) is the existence of a percolation threshold $p=p_c$ \cite{DIM_Stauffer}.
For $p=1$, the system is pure and shows ferromagnetic order at $T<T_c(p=1)$.
The critical temperature $T_c(p)$ diminishes as $p$ is decreased and becomes
0 at $p=p_c$. (For a $d=2$ square lattice, $p_c \simeq 0.593$.) For $p<p_c$,
there are no infinite clusters of magnetic atoms which span the system, i.e.,
there is no long-range order. For weak disorder ($p \simeq 1$), the kinetic
DIM is analogous to the kinetic RBIM. However, for smaller values of $p$,
connectivity effects become important and may change the nature of domain
growth. We are particularly interested in the ordering dynamics of the DIM for
$p\simeq p_c$.

In this section, we focus on two systems: \\
(a) The DIM with nonconserved (Glauber) kinetics, which models the
ordering dynamics of a dilute ferromagnet. In an MC simulation of Glauber
kinetics, a randomly-chosen spin $S_i$ is flipped to $-S_i$ and the system is
evolved according to the prescription in Eq.~(\ref{tp}). \\
(b) The DIM with conserved (Kawasaki) kinetics, which models the
segregation kinetics of a dilute binary mixture. In this case,
we use the continuous-time algorithm described in Sec~III.A.

The initial conditions for our MC simulations are prepared as follows. We
dilute the sites of an $L^2$ lattice with probability $1-p$. (These sites remain
fixed during the evolution.) Then, up and down spins are randomly distributed
on the remaining sites with a zero net magnetization, mimicking the
high-temperature disordered configuration before the quench.

\subsection{Nonconserved Kinetics}

\begin{figure}[htb]
\centering
\includegraphics[width=\linewidth]{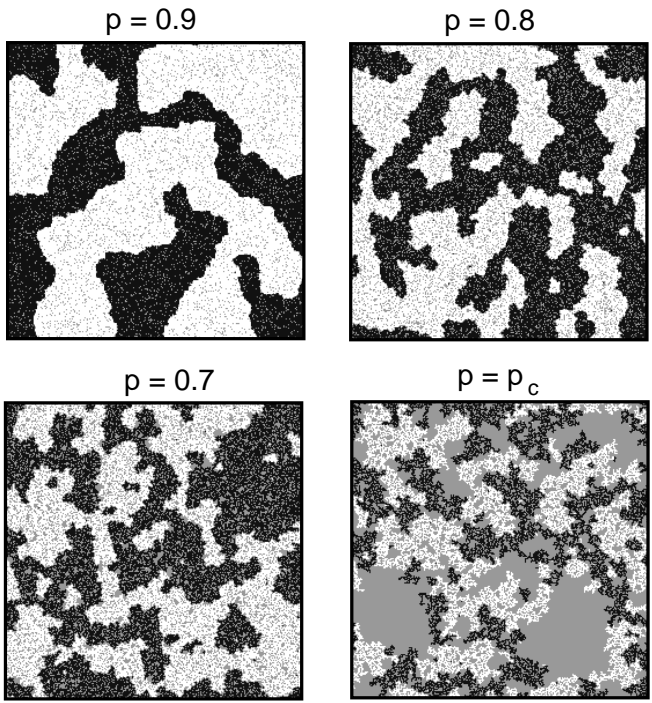}
\caption{Domain growth in the DIM with Glauber kinetics. We show evolution
pictures at $t=10^{6}$ MCS for a $256^2$ corner of a $512^2$ lattice,
after a quench from $T = \infty$ to $T = 0.5$. The snapshots correspond to
different site occupation probabilities: $p= 0.9, 0.8, 0.7, 0.593~(p_c)$.
The up and down spins are marked black and white, respectively. The
missing spins are marked grey.}
\label{fig7}
\end{figure}

In Fig.~\ref{fig7}, we show evolution snapshots at $t=10^{6}$ MCS for $T=0.5$
and $p=0.9, 0.8, 0.7, 0.593~(p_c)$.
Notice that $T_c(p=0.7) \simeq 1.04$
for the $d=2$ DIM \cite{ch76},
so that $T=0.5$ lies below the critical temperature for all
the values of $p$ other than $p=p_c$, where $T_c(p_c)=0$. (Unfortunately, it
is difficult to do MC simulations at $T=0$, as the system is rapidly
trapped in a metastable state.) As expected, the domain size at a
fixed time diminishes with increase in disorder. In the case of evolution
on the backbone of a percolating cluster, the morphology consists
of a network of islands (compact well-connected regions) linked by just a
single bond. As time progresses, these islands become fully magnetized, but they
cannot influence the evolution of their neighbors. For this reason,
domain growth becomes very slow at $p=p_c$. Further, as $T >T_c(p_c)$, domain
growth is arrested when the length scale saturates at the equilibrium correlation
length $\xi_{\rm{eq}} (T) \rightarrow \infty$ as $T \rightarrow 0$.

\begin{figure}[htb]
\centering
\includegraphics[width=\linewidth]{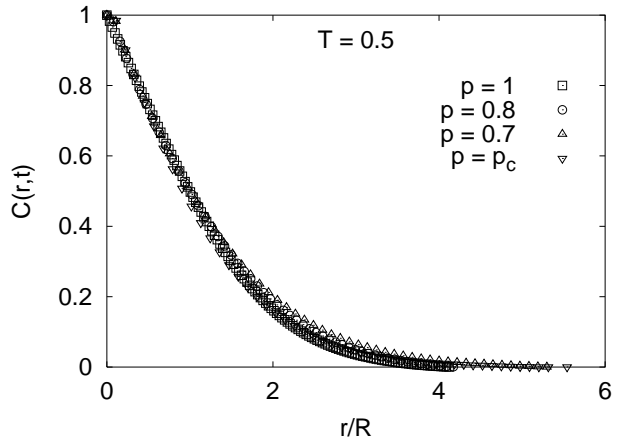}
\caption{Scaling plot of the correlation function for the evolution
depicted in Fig.~\ref{fig7}. We plot $C(r,t)$ vs.
$r/R$ at $t=10^6$ MCS for occupation probability $p = 0.8,0.7,p_c$.
We also show data for the pure case ($p=1$) at $t=10^3$ MCS, obtained
for a $1024^2$ system.}
\label{fig8}
\end{figure}

Let us next focus on the properties of these evolution
morphologies. Our statistical data for the nonconserved DIM
is obtained using $512^2$ systems, by averaging
over 50 independent initial conditions and disorder configurations.
We have confirmed that the evolution of the nonconserved DIM shows
dynamical scaling. In Fig.~\ref{fig8}, we demonstrate the
disorder-independence of the scaled correlation function. Here,
we plot $C(r,t)$ vs. $r/R$ at $t=10^6$ MCS for $p = 0.8,0.7,p_c$,
and compare it with the corresponding data for the pure case ($p=1$).
In this case, the domain size is defined as the $r$-value
where the correlation function decays to half its maximum value. Notice
that the scaling function for $p=p_c$ is analogous to that
for higher values of $p$, and there are no distinctive signatures
of the percolation cluster.

\begin{figure}[htb]
\centering
\includegraphics[width=\linewidth]{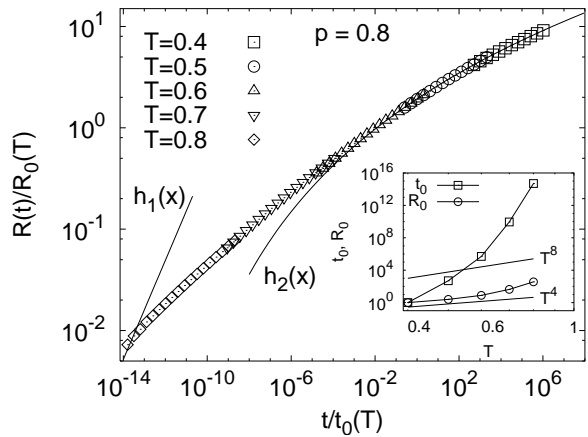}
\caption{Scaling plot to test the crossover function in
Eqs.~(\ref{cross})-(\ref{hx}). For each
temperature $T$, the values for $R_0(T)$ and $t_0(T)$ have been chosen
to obtain a smooth scaling curve $h(x)$. The
functions $h_1(x) \propto x^{1/2}$ and $h_2(x) \propto (\ln x)^4$ represent
the expected asymptotic behavior for $x\ll1$ and $x\gg1$,
respectively. The inset shows the temperature-dependence of the fit
values $R_0(T)$ and $t_0(T)$, and their expected $T$-dependence,
which is $T^4$ and $T^8$, respectively.}
\label{fig9}
\end{figure}

Next, consider the time-dependence of the length scale. We first study
the case with $p=0.8$ and varying $T$-values. In Fig.~\ref{fig9}, we attempt
to fit our length-scale data to the HH crossover function in
Eqs.~(\ref{cross})-(\ref{hx}). We record the following points of
disagreement with the HH scaling behavior: \\
(a) The short-time behavior is not described well by Eq.~(\ref{hx}),
where $h_1(x) \sim x^{1/2}$. \\
(b) The asymptotic behavior in Eq.~(\ref{hx})
[denoted by the curve $h_2(x)$ in Fig.~\ref{fig9}] does not fit
the scaling curve well even for the largest times. \\
(c) The temperature-dependence of the crossover
length $R_0(T)$ and the crossover time $t_0(T)$ is stronger than a
power law (see inset), which is incompatible with Eqs.~(\ref{hh}) and
(\ref{r0}). The parameter $a_0$ in Eq.~(\ref{hh}) is proportional to
the surface tension, and is expected to decrease with increasing
temperature. Therefore, $t_0$ can be expected to increase
faster than $T^8$, but its $T$-dependence turns out to
be much too strong: note 
that $t_0$ in the inset of Fig.~\ref{fig9} varies over 20
decades when $T$ varies over only half a decade from 0.4 to 0.8.
We do not see why the surface tension should have such a strong
$T$-dependence. \\
Based on observations (a)-(c), we believe that 
the data in Fig.~\ref{fig9} is inconsistent with the HH growth law.
\begin{figure}[htb]
\centering
\includegraphics[width=\linewidth]{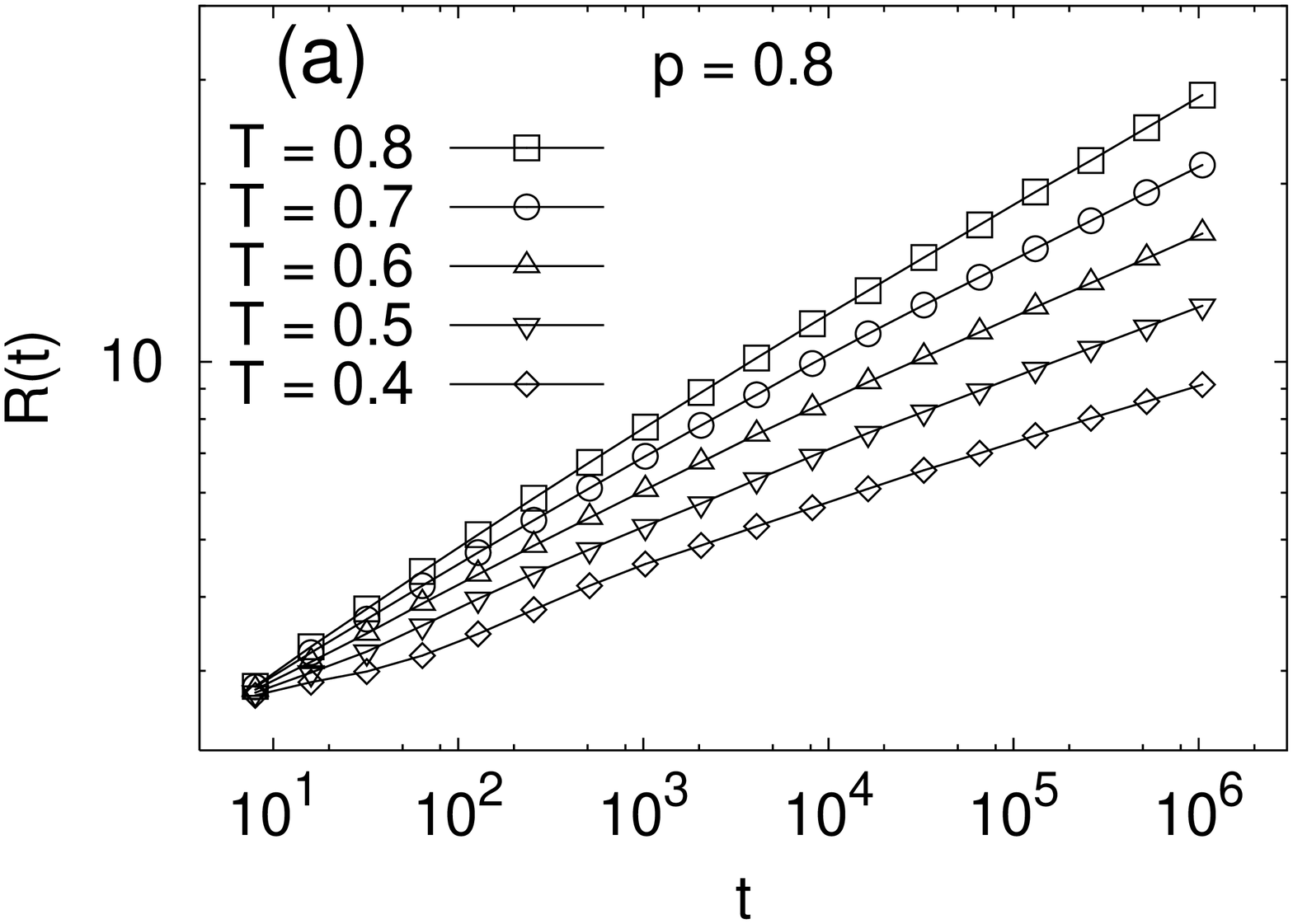}
\includegraphics[width=\linewidth]{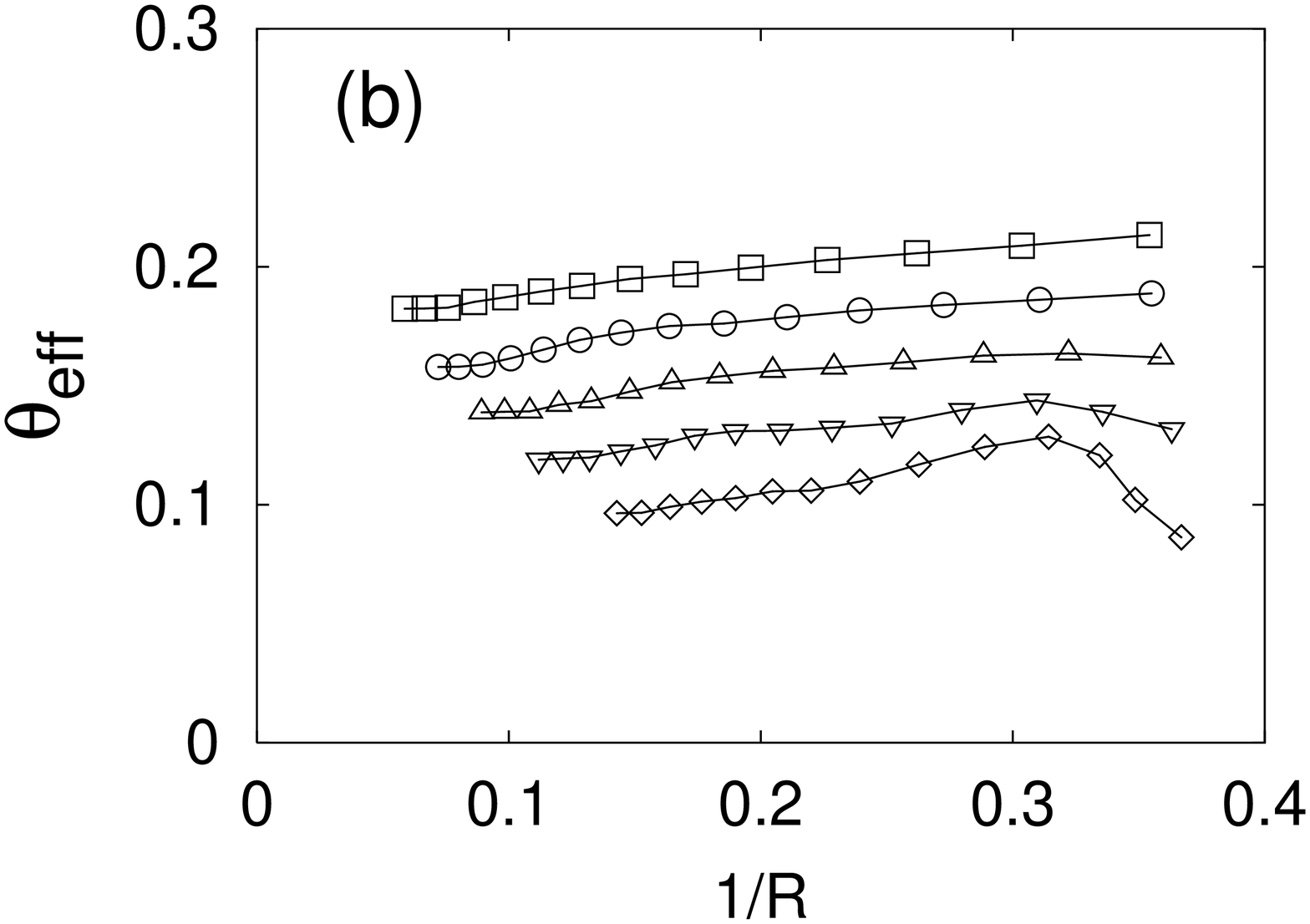}
\includegraphics[width=\linewidth]{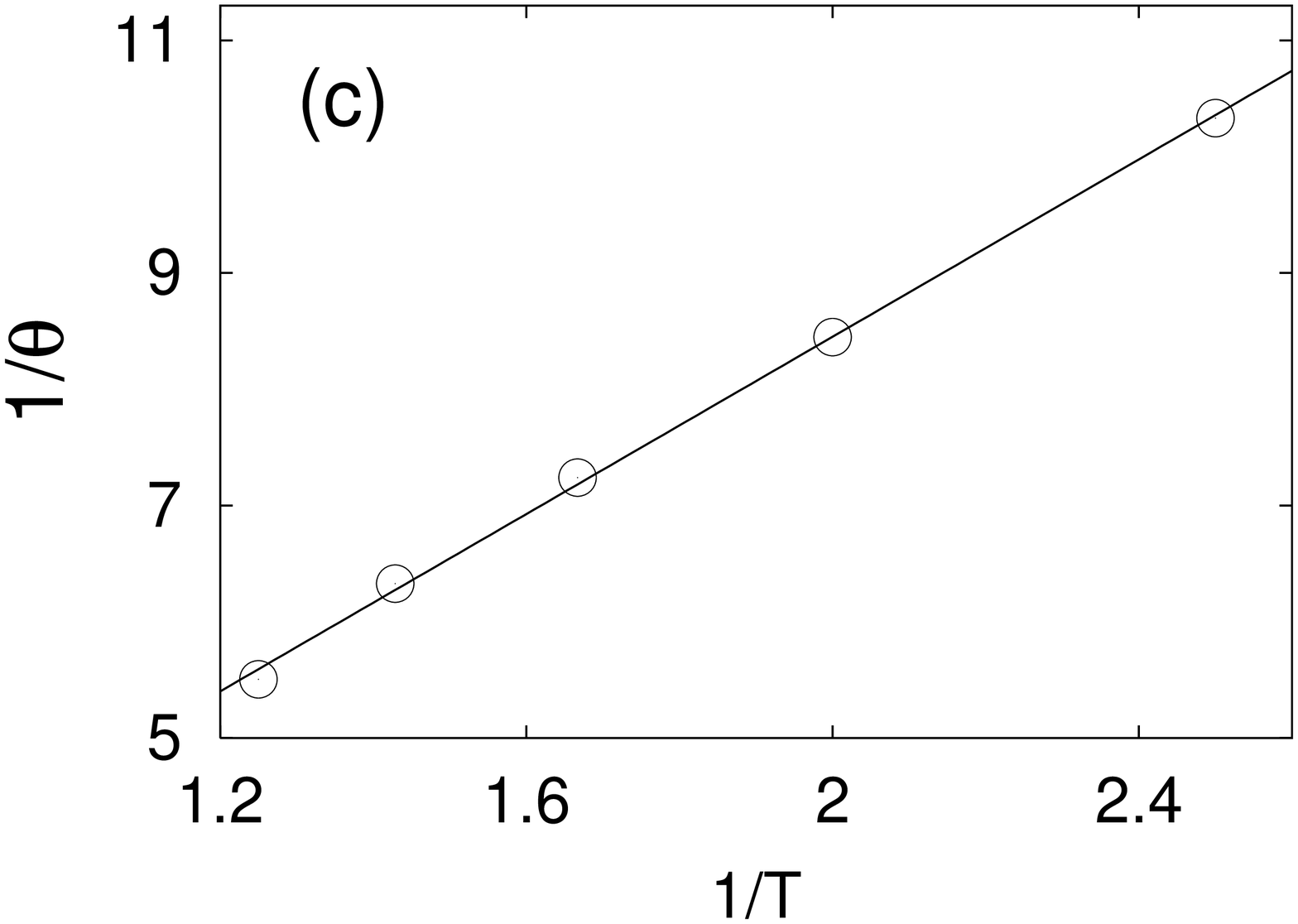}
\caption{(a) Plot of $R$ vs. $t$ (on a log-log scale) for the nonconserved
DIM with $p = 0.8$ and temperatures $T = 0.4,0.5,0.6,0.7,0.8$. (b) Plot of
$\theta_{\rm{eff}} = d(\ln R)/d(\ln t)$ vs. $R^{-1}$ for the data in (a).
(c) Plot of $\theta^{-1}$ vs. $T^{-1}$ for the data in (a).}
\label{fig10}
\end{figure}
In Fig.~\ref{fig10}(a), we plot $R$ vs. $t$ on a log-log scale
for $p = 0.8$ and $T=0.8,0.7,0.6,0.5,0.4 < T_c(p=0.8) \simeq 1.5$. The
corresponding plots of $\theta_{\rm{eff}}$ vs. $R^{-1}$ are shown in
Fig.~\ref{fig10}(b). These show an extended flat regime, making it
relatively simple to estimate the exponent. As in the case of the RBIM,
our data is consistent with power-law growth with a variable
exponent. In Fig.~\ref{fig10}(c), we plot $\theta (T,p)^{-1}$ vs. $T^{-1}$ --
the linear behavior is consistent with Eq.~(\ref{ncexp}). (See Ref.~\cite{ppr04}
for similar results for the nonconserved RBIM.)

\begin{figure}[htb]
\centering
\includegraphics[width=\linewidth]{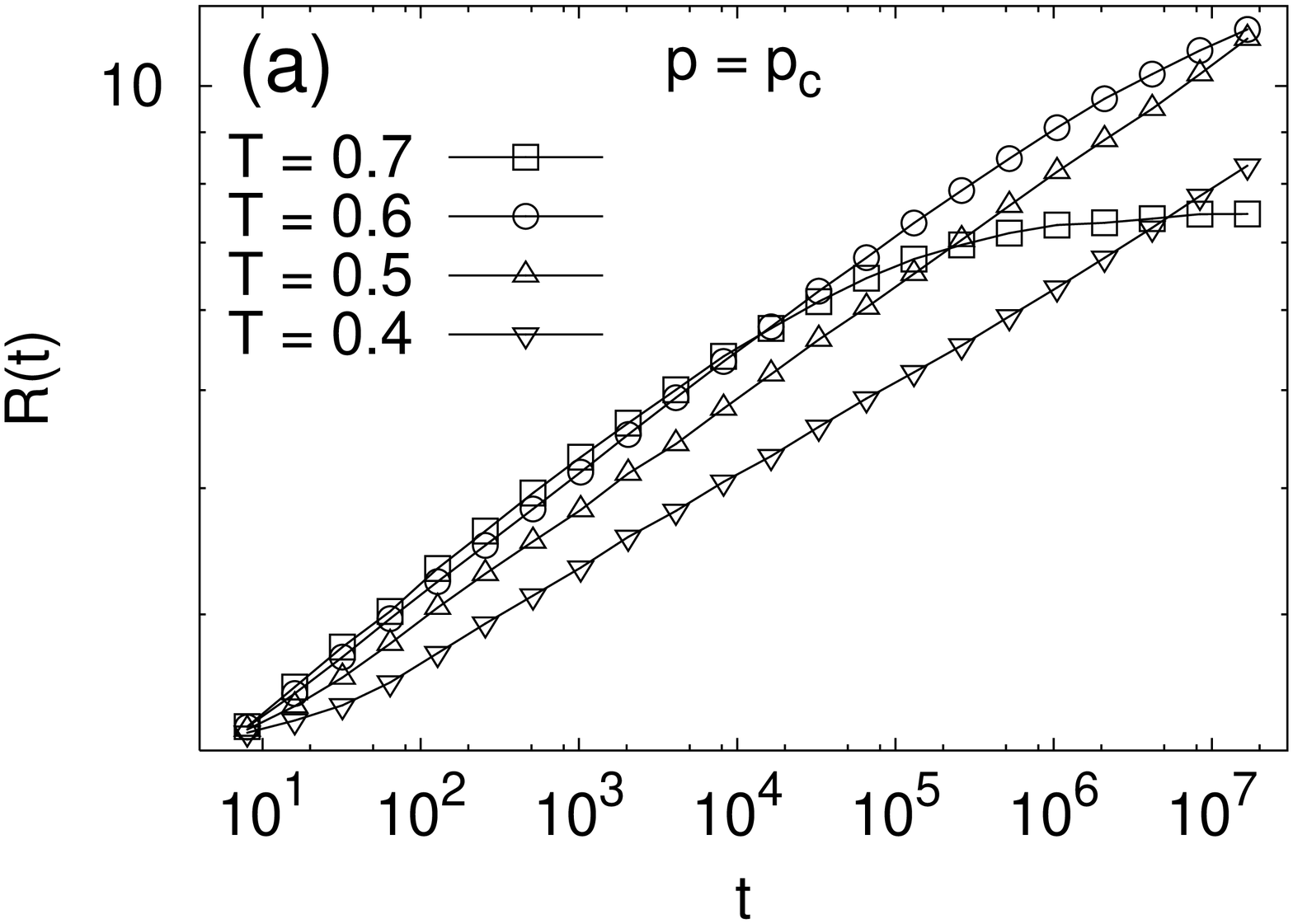}
\includegraphics[width=\linewidth]{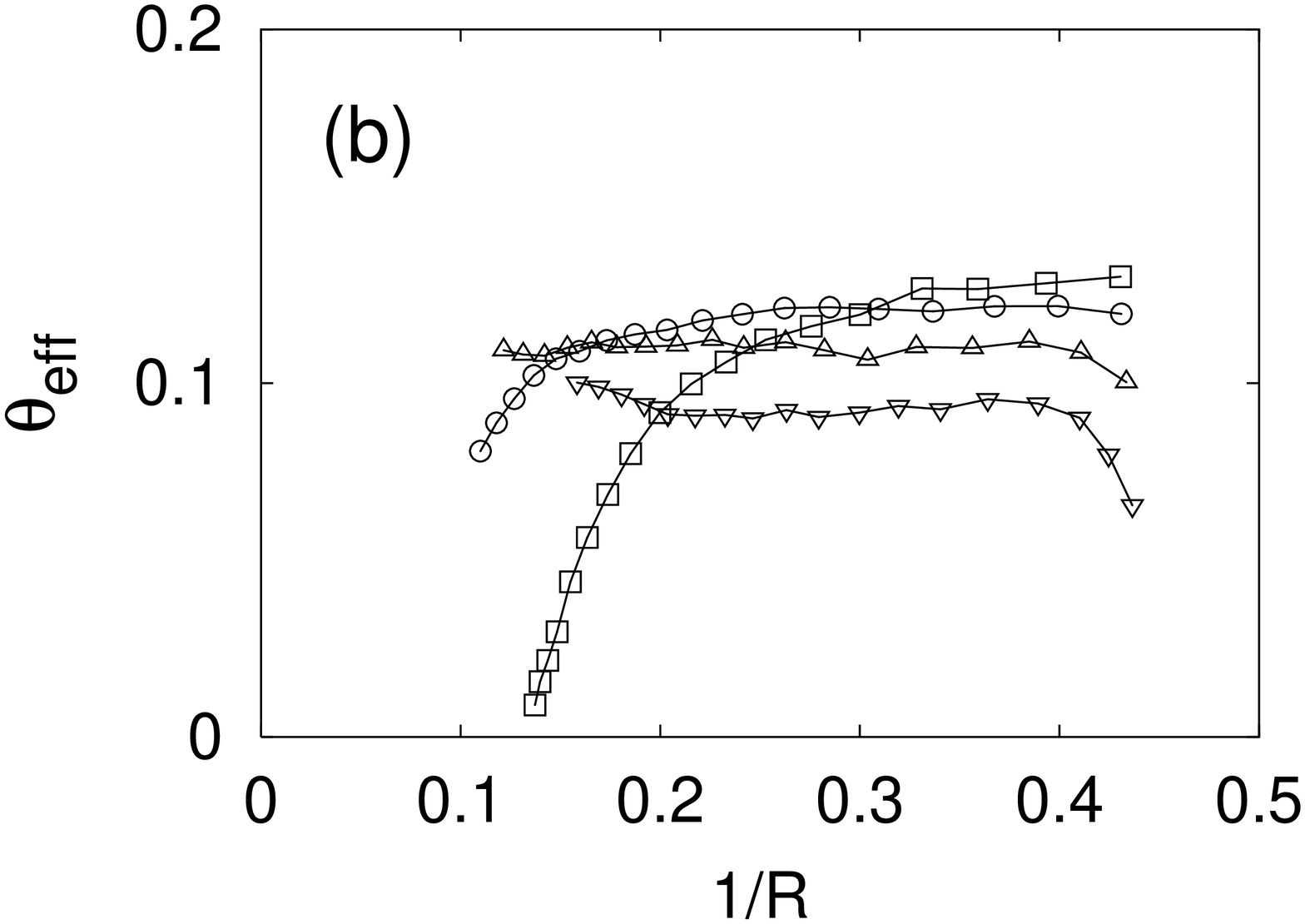}
\includegraphics[width=\linewidth]{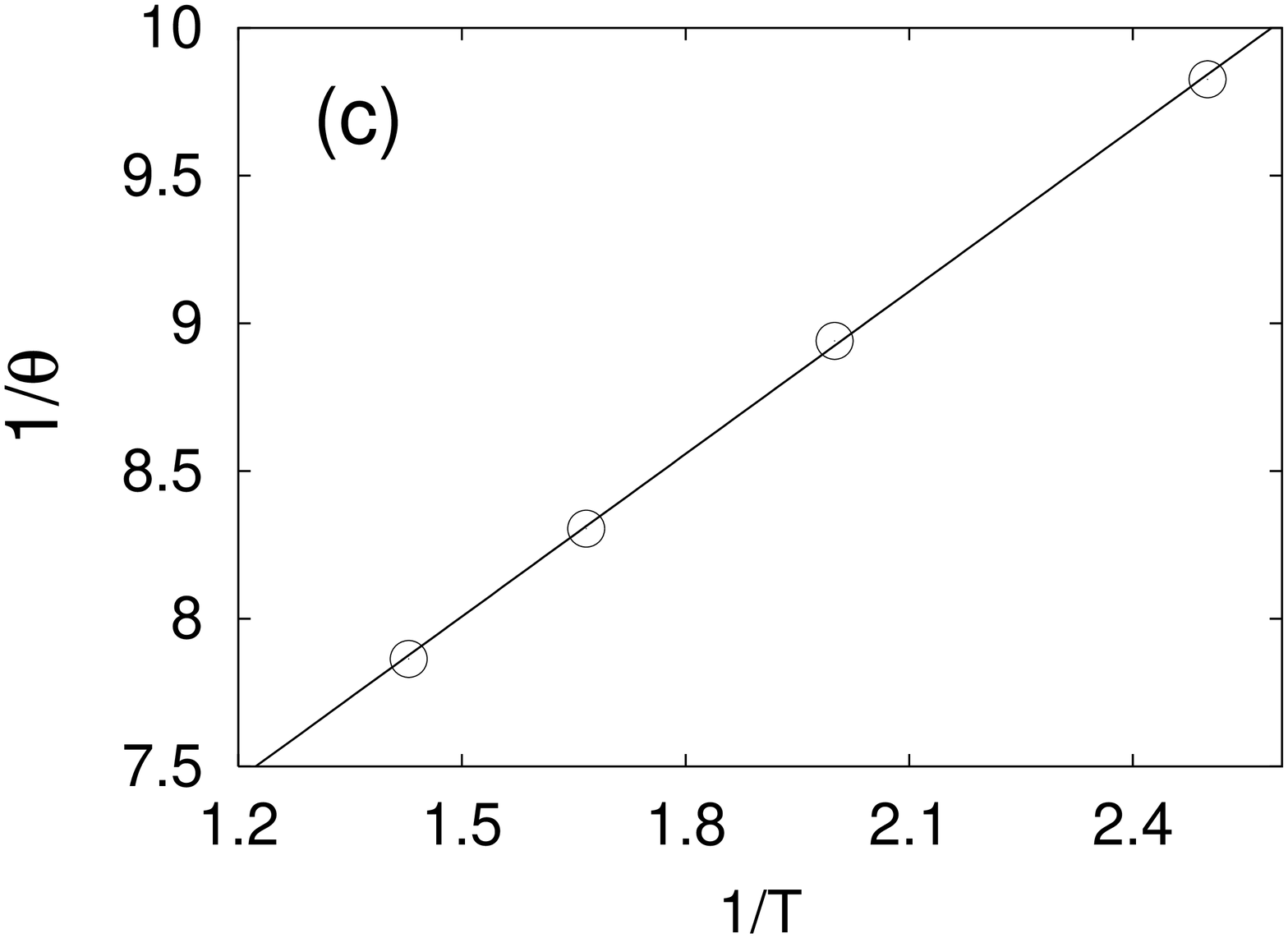}
\caption{Analogous to Fig.~\ref{fig10} but for $p = p_c$.}
\label{fig11}
\end{figure}
Finally, in Fig.~\ref{fig11}(a), we plot $R$ vs. $t$ at percolation
$(p=p_c)$ and $T=0.7,0.6,0.5,0.4 > T_c(p_c)=0$. 
Recall that the domain scale saturates to $\xi_{\rm{eq}}(T)$ in this case, with an
earlier crossover for higher $T$. On the time-scale of our simulation, the data
for $T=0.7$ has saturated, and that for $T=0.6$ is beginning to bend over. This is
reflected in Fig.~\ref{fig11}(b), which shows $\theta_{\rm{eff}}$ vs. $R^{-1}$.
The exponent $\theta$ is estimated from the flat portion of these curves, and
we plot $\theta^{-1}$ vs. $T^{-1}$ in Fig.~\ref{fig11}(c).

\subsection{Conserved Kinetics}

We have performed a similar study of the DIM with Kawasaki kinetics. In this
case, the time-scale of growth is considerably slower than for the nonconserved
case.
\begin{figure}[htb]
\centering
\includegraphics[width=\linewidth]{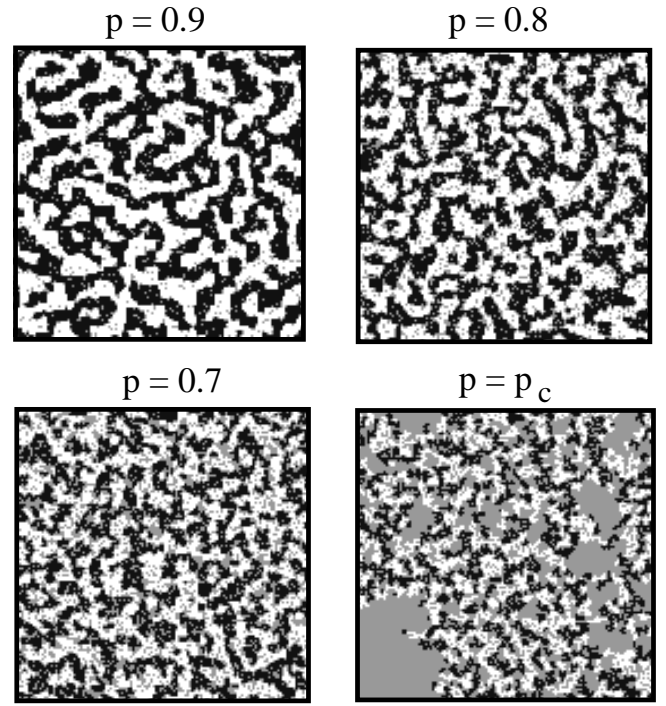}
\caption{Domain growth in the DIM with Kawasaki kinetics. We show evolution
pictures at $t=10^{7}$ MCS for a $128^2$ corner of a $256^2$ lattice,
after a quench from $T = \infty$ to $T = 0.5$. The snapshots correspond
to different site occupation probabilities: $p= 0.9, 0.8, 0.7, 0.593~(p_c)$.
The color coding is the same as in Fig.~\ref{fig7}.}
\label{fig12}
\end{figure}
The typical evolution morphologies at $t = 10^7$ MCS (after a critical quench
from $T=\infty$ to $T=0.5$) are shown in Fig.~\ref{fig12}. As in the
earlier cases, we will show results for the correlation function and
the growth law. 
The statistical data shown here was obtained on a $256^2$ lattice as an average
over 32 independent configurations.

\begin{figure}[htb]
\centering
\includegraphics[width=\linewidth]{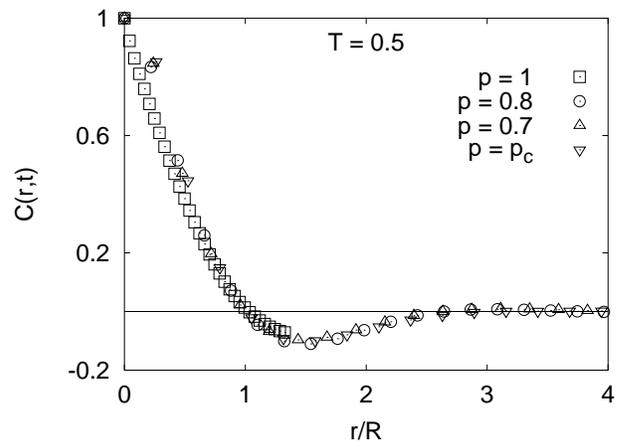}
\caption{Scaling plot of the correlation function for the
evolution depicted in Fig.~\ref{fig12}. We
plot $C(r,t)$ vs. $r/R$ at $t=10^7$ MCS for occupation probability
$p = 1$ (pure case) and $0.8,0.7,p_c$.}
\label{fig13}
\end{figure}

In Fig.~\ref{fig13}, we plot $C(r,t)$ vs. $r/R$ at $t=10^7$ MCS for the pure
case, and different values of the dilution. [The length scale is obtained
from the first zero of $C(r,t)$.] Again, the scaling function is approximately
independent of the amount of dilution. Next, we focus on the time-dependence
of the length scale. 
\begin{figure}[htb]
\centering
\includegraphics[width=\linewidth]{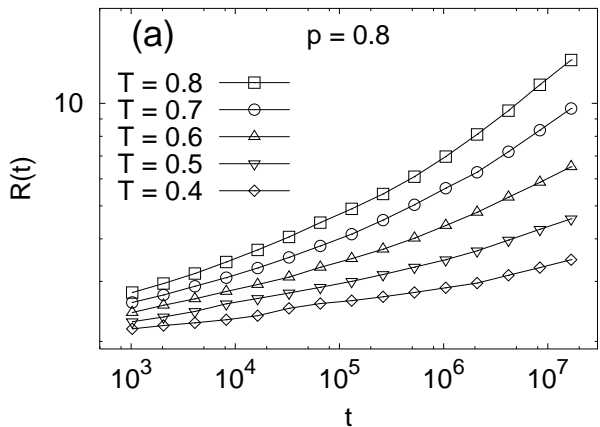}
\includegraphics[width=\linewidth]{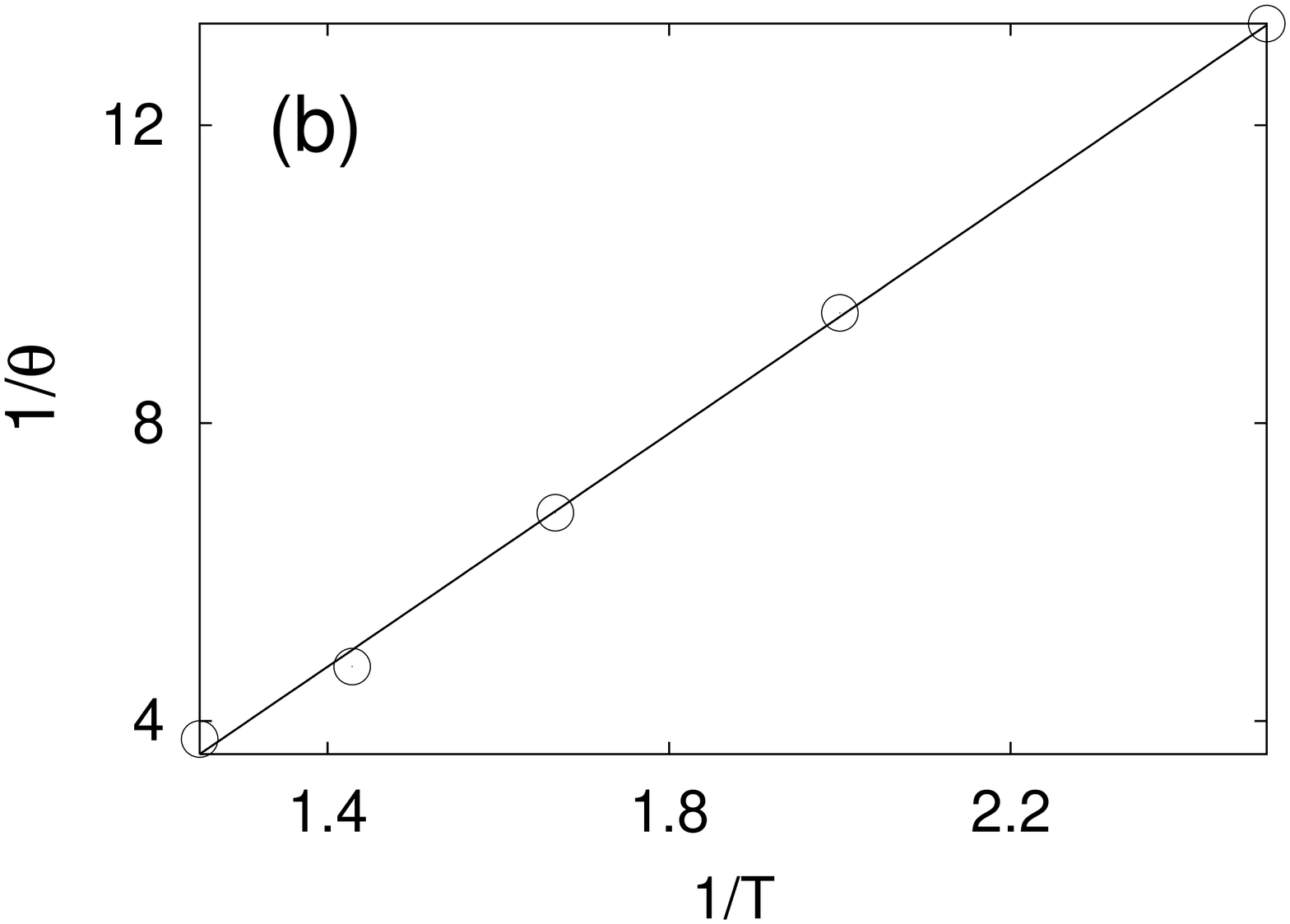}
\caption{(a) Plot of $R$ vs. $t$ (on a log-log scale) for the conserved
DIM with $p = 0.8$ and $T = 0.4,0.5,0.6,0.7,0.8$. (b) Plot of $\theta^{-1}$
vs. $T^{-1}$ for the data in (a).}
\label{fig14}
\end{figure}
In Fig.~\ref{fig14}(a), we plot $R$ vs. $t$ for $p=0.8$
and various values of $T$. Again, we estimate the asymptotic exponent from
plots of $\theta_{\rm{eff}}$ vs. $R^{-1}$ (not shown here). 
In Fig.~\ref{fig14}(b), we plot the corresponding $\theta^{-1}$ vs. $T^{-1}$.

\begin{figure}[htb]
\centering
\includegraphics[width=\linewidth]{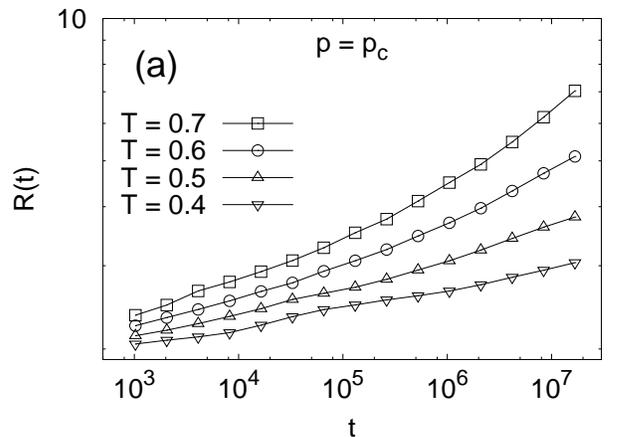}
\includegraphics[width=\linewidth]{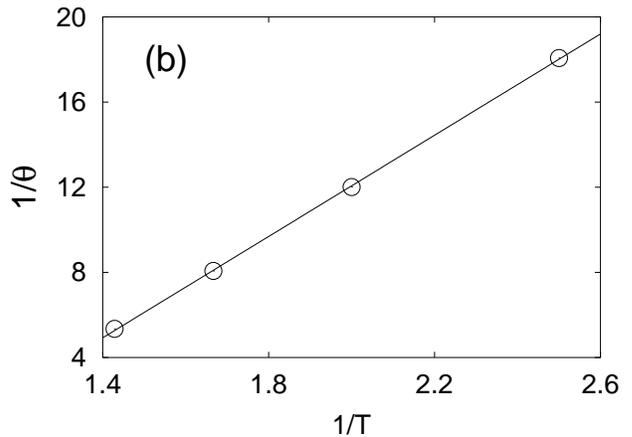}
\caption{Analogous to Fig.~\ref{fig14}, but for $p=p_c$.}
\label{fig15}
\end{figure}
Figure~\ref{fig15} is analogous to Fig.~\ref{fig14}, but for $p=p_c$.
As the growth is much slower than the nonconserved case, we do not
see a crossover to saturation for $p = p_c$ on the time-scale of our
simulations. Once again, the exponents are consistent with the logarithmic
barrier-scaling scenario.

\section{Summary and Discussion}

Let us conclude this paper with a summary and discussion of the results
presented here and in our earlier letter \cite{ppr04}. We have undertaken
comprehensive Monte Carlo (MC) simulations of domain growth in Ising
systems with quenched disorder. These studies are based on kinetic
Ising models with either nonconserved (Glauber) spin-flip kinetics
or conserved (Kawasaki) spin-exchange kinetics.
The nonconserved case models ordering dynamics
in random magnets, and the conserved case models segregation kinetics in
disordered binary mixtures. We have studied
domain growth for two classes of disordered systems: \\
(a) The {\it random-bond Ising model} (RBIM), where the exchange interaction
has a uniform distribution on the interval $[1-\epsilon/2, 1+\epsilon/2]$, 
$\epsilon < 2$. In this case, the critical temperature $T_c(\epsilon)$
remains approximately unchanged. \\
(b) The {\it dilute Ising model} (DIM), where the exchange interaction is
randomized by the dilution of magnetic atoms with non-magnetic impurities.
In this case, the critical temperature $T_c(p)$ ranges from $T_c(p=1)
\simeq 2.269$ (in $d=2$) to $T_c(p=p_c) = 0$ ($p_c \simeq 0.593$ in $d=2$). \\
Both classes of disorder are of considerable experimental relevance.

The general framework for understanding coarsening in disordered
systems is as follows. At early times, the domain sizes are small and
domain growth is unaffected by disorder. At late times, the domain
boundaries are trapped by disorder sites, and asymptotic growth
proceeds via thermally-activated hopping over disorder barriers.
Clearly, the asymptotic growth law depends critically on the
length-scale dependence of the disorder barrier $E_B$.
In this context, an important study is due to Huse
and Henley (HH) \cite{hh85}. In the HH scenario, the disorder
barriers have a power-law dependence on the
domain size, $E_B \sim R^\psi$. These result in a logarithmic
domain growth law in the asymptotic regime.
We find that our MC results are not in agreement with the HH scenario.
Rather, our results are consistent with power-law growth
with an exponent $\theta$ which depends on the temperature $T$ and
the disorder amplitude $\epsilon$. This is in agreement with a number
of experiments \cite{iei90,lla00,llop01}, and early simulations of droplet
shrinking in disordered systems by Oh and Choi \cite{oc86}. This scenario
arises naturally in the context of logarithmic energy barriers, and the
corresponding functional dependence of $\theta (T,\epsilon)$
is in excellent agreement with our numerical results.

Our results provide a framework for the analysis of experiments and simulations
on domain growth in disordered magnets and binary mixtures. We hope that
our study will motivate fresh experimental studies of this important
problem. In particular, there is a paucity of experimental results on
phase separation in disordered mixtures.

Acknowledgement: This work was financially supported by the Deutsche
Forschungsgemeinschaft~(DFG), SFB277.


\begin{thebibliography}{99}
\bibitem{XX} Variation of $T_c$ for RBIM has been estimated from
        Binder Cumulant.
\end{thebibliography}
\end{document}